\documentclass[sigconf]{acmart}
\usepackage{float, svg, graphicx, multirow}

\copyrightyear{2019} 
\acmYear{2019} 
\setcopyright{acmcopyright}
\acmConference[SoICT 2019]{The Tenth International Symposium on Information and Communication Technology}{December 4--6, 2019}{Hanoi - Ha Long Bay, Vietnam, Viet Nam}
\acmBooktitle{The Tenth International Symposium on Information and Communication Technology (SoICT 2019), December 4--6, 2019, Hanoi - Ha Long Bay, Vietnam, Viet Nam}
\acmPrice{15.00}
\acmDOI{10.1145/3368926.3369700}
\acmISBN{978-1-4503-7245-9/19/12}

\acmSubmissionID{89}

\begin{document}

\title{Deep Learning Approach for Singer Voice Classification of Vietnamese Popular Music}

\author{Toan Pham Van}
\affiliation{%
  \institution{R\&D Lab, Sun* Inc}
  \city{Hanoi}
  \country{Vietnam}
}
\email{pham.van.toan@sun-asterisk.com}

\author{Ngoc Tran Ngo Quang}
\affiliation{%
  \institution{R\&D Lab, Sun* Inc}
  \city{Hanoi}
  \country{Vietnam}
}
\email{tran.ngo.quang.ngoc@sun-asterisk.com}

\author{Ta Minh Thanh}
\affiliation{%
  \institution{Le Quy Don Technical University}
  \streetaddress{236 Hoang Quoc Viet}
  \city{Hanoi}
  \country{Vietnam}}
\email{thanhtm@mta.edu.vn}

\renewcommand{\shortauthors}{Toan et al.}

\begin{abstract}
Singer voice classification is a meaningful task in the digital era. With a huge number of songs today, identifying a singer is very helpful for music information retrieval, music properties indexing, and so on. In this paper, we propose a new method to identify the singer's name based on analysis of Vietnamese popular music. We employ the use of vocal segment detection and singing voice separation as the pre-processing steps. The purpose of these steps is to extract the singer's voice from the mixture sound. In order to build a singer classifier, we propose a neural network architecture working with Mel Frequency Cepstral Coefficient (MFCC) as extracted input features from said vocal. To verify the accuracy of our methods, we evaluate on a dataset of 300 Vietnamese songs from 18 famous singers. We achieve an accuracy of 92.84\% with 5-fold stratified cross-validation, the best result compared to other methods on the same data set.
\end{abstract}



\maketitle

\section{Introduction}
\subsection{Overview}
With the growing collections of digital music, the number of songs are published increases day by day. As a result, we need an automatic system that can classify each song with useful information such as singer or categories. This task would be useful for music retrieval problems, automatic database indexing, or content-based music recommendation systems \cite{content_based}. It also is important to contribute to future research on music index retrieval (MIR) for Vietnamese songs. Each singer's vocal has different acoustic features as timbre, pitch, frequency range; and all of them are useful for singer classification. From these features, we can use some machine learning method to classify the singer's gender or age \cite{singer_gender_age}, or timbre \cite{singer_timbre}. Similarly, our problem could also be solved with the approach of machine learning. Specifically, from the audio characteristics of vocals, we may use traditional algorithms such as Support Vector Machine (SVM), Naive Bayes, or k-Nearest Neighbor (KNN), to classify the singers. On another note, the application of deep learning methods has been very popular lately: It is used in most fields of artificial intelligence such as computer vision, natural language processing, audio processing, etc \cite{deep_learning}. The application of this technology to the singer classification problem is a new approach. In light of this, we propose a new scheme for singer classification in this paper using cutting-edge deep neural network models to achieve state-of-the-art result, exceeding which of classical methods. We divide this process into 3 phases: vocal segmentation, vocal extraction, and vocal classification. The overall architecture of our system is demonstrated in \textbf{Fig~\ref{overview}}. Each of these steps is solved by a different neural network architecture trained on different datasets. We have improved on each of these components with our customized neural network architectures to help the overall system achieving better accuracy.
\subsection{Our contributions}
Our contributions are as follows:

\begin{itemize}
  \item We propose a new method to identify the singers based on their song. This is a new method in music index retrieval (MIR) research field.
  \item We design a new neural network architecture (NNA) to achieve the better accuracy in each sub-problem. According to the proposed NNA, our paper achieves an overall result superior to which from using traditional methods such as SVM (support vector machine), KNN (k-nearest neighbor algorithm), Naive Bayes, and so on.
  \item We build a dataset which can be used publicly available for research purposes based on Vietnamese popular songs. Also, our dataset is published for non-commercial applications. 
\end{itemize}
\subsection{Roadmap}
The rest of this paper is organized as follows. Section 2 presents a brief review of related works. All the deep learning methods used to identify singers are mentioned in Section 3. Section 4 describes the system setup for the experiments and our dataset. Experimental results and evaluations are presented in Section 5. Finally, we conclude the paper in Section 6.

\begin{figure*}[tb]
    \centering
    \includegraphics[width=14cm]{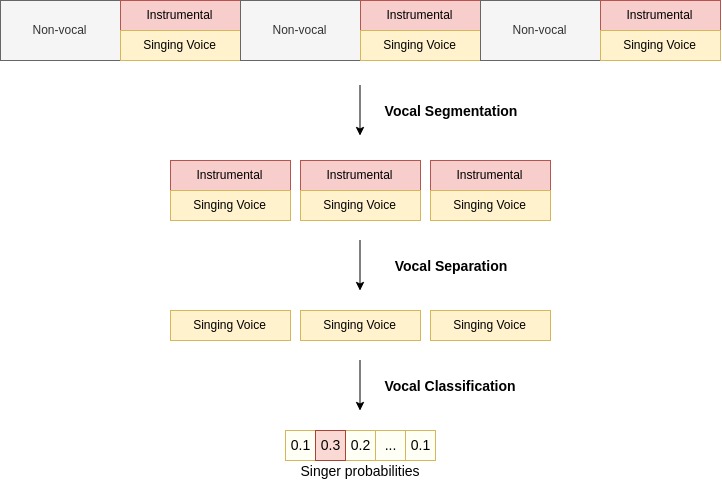}
    \caption{The overview of our singer classification system.}
    \label{overview}
\end{figure*}

\section{Related works}
\subsection{Overview of problems}
Nowadays, in order to meet the demands of human entertainment, music works are produced with increasing frequency. The demand for searching and storing information related to songs has also increased accordingly. Music databases are rapidly expanding as a result, storing in millions of tracks in the digital cloud. Meanwhile, there is a lack of meta-information for a majority of them. The problem of automatically classifying music information ($e.g.$ singer, music genres, etc) becomes very meaningful. This is also a research topic that deserves attention in the field of computer science. The purpose of our research is to predict the singer with the highest accuracy from any short song snippet. Our classification method can be applied on the music index retrieval, music searching system, and music categorization. The mentioned problem of automatically classifying music can be separated into the following steps:

\begin{itemize}
\item\textbf{Step 1 - Vocal Segmentation:} Main purpose of this task is splitting the input audio into vocal and non-vocal (instrumental) regions. This gives prior information that the audio passing though the next step which is vocal separation, has a vocal component.

\item\textbf{Step 2 - Vocal Separation:} After distinguishing between vocal and non-vocal segments from the whole song, we proceed to vocal separation. This step extracts the vocal (singers' voice) from the mixture sound mixing together drums, guitar, background music, and so on. This is an important phase to determine the accuracy of the singer classification later. The clearly vocal and non-vocal separation can be make the clean features for preparing the input data of last step. 

\item\textbf{Step 3 - Vocal Classification:} After obtaining a clean vocal ($i.e.$ removed most of the noise), we continue on to use a neural network to classify singers. This problem is quite similar to the speaker identification but for singing vocal instead. In this step, the feature of singers' voice can be clearly classified using our proposed method.  
\end{itemize}

\subsection{Audio acoustic features}
For any machine learning model to work properly and effectively, good feature extraction is required. We employ two standard feature generating operations in audio processing such as the Short-Time Fourier Transform (STFT) \cite{time-freq_feat_rep} and the Mel-Frequency Cepstrum (MFC) \cite{mfc}. 

STFT provides us the frequencies' properties in short timeframes and describes how it changes over time. On the other hand, the MFC feature makes use of the magnitude of those information, maps it into a more natural-sounding frequency domain (the Mel scale), and keeps only informative parts of those by discarding some coefficients. 

Existing literature provides a plethora of fingerprinting methods from spectrograms \cite{fingerprinting}. In this paper, we opt to extract deeper features from such existing schemes through the use of deep neural networks (DNN). STFT and MFCC features are employed for efficiently audio acoustic features extraction.

\begin{figure*}[t]
    \centering
    \includegraphics[width=0.88\textwidth,height=4.5cm]{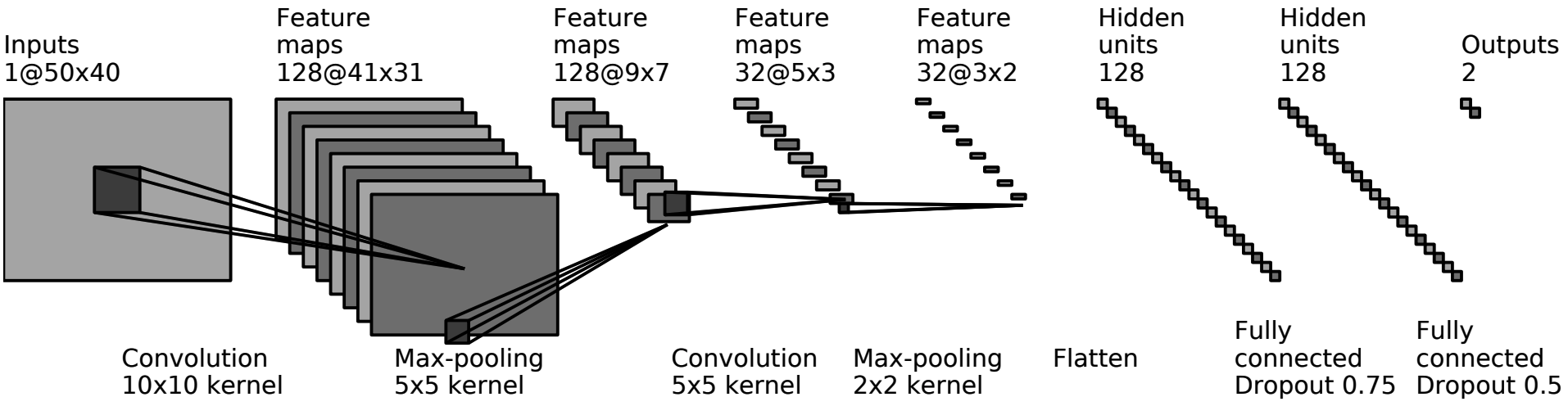}
    \caption{The architecture of our vocal/non-vocal segmentation model using Convolutional Neural Network.}
    \label{cnn_figure}
\end{figure*}

\subsection{Vocal segmentation problem}
In any song, some segments of a song are purely instrumental ($e.g.$, intro, bridge, and outro) and these are not needed for the vocal classification task. Instead of processing with the whole song, it makes sense to ignore these non-vocal segments via a vocal segmentation algorithm. There have been numerous studies and methods in this field, such as formant-based \cite{formant_segment}, frequency-based \cite{freq_response}, Hidden Markov model \cite{hmm_segment}, etc. In this work, we use deep convolutional neural network with input audio features.

\subsection{Vocal separation problem}
Before singers classification, we need to extract the vocal-only parts from a master mix. There has been a proliferation of literature in the subject from the music information retrieval and the singing voice separation communities; since beyond our use, this topic is worth billions in the entertainment industry. Classical methods would use techniques like variants of non-negative matrix factorization (NMF) \cite{nmf_vox}, where one can think of each component in the decomposition being an instrumental track. More modern researches would use ideal/soft binary mask on the spectrogram of the master mix, using either image segmentation methods like U-Net \cite{spotify-unet}, or convolutional neural networks followed by a simple flatten-dense layer.

\subsection{Singer classification problem}
And last but not least is our final step to output the singer from the extracted vocal track. Previous works in speaker detection used classical methods like Gaussian Mixture Models \cite{ira_dataset_2}, while newer ones employ modern convolutional neural networks \cite{speaker-cnn}. For singer detection, the paper \cite{freq_response} has used acoustic features like frequency responses to achieve some decent result. Like every other classification problems, after extracting deep features, such features are fed into some differentiator such as multiclass SVM or a densely-connected network with softmax activation.

During the process of finalizing this paper, we came across some new works on this topic of singer classification \cite{singer-chinese}. The main contribution of \cite{singer-chinese} is to propose a new deep learning approach based on LSTM and MFCC features to identify the singer of a song in large datasets. In another work related with singer classification problem, the method in paper \cite{mfcc-lpc} employed MFCC and LPC (linear predictive coding) coefficients from Indian video songs as the singers' feature, and then the singer models are trained using Naive bayes classifier and back propagation algorithm using neural network. Both of related works seemed to try separating the vocal and non-vocal from background soundtrack. However, their pre-process is not considered carefully. 

While the actual configurations of the network contain notable differences, the general idea behind the architecture design is relatively similar. We have plan on comparing the models and analyzing the effects of those differences, but that will not be in the scope of this paper.

\section{Deep learning algorithms}
In this section, we present the deep learning algorithms used to solve our defined problem above. With each task in workflow shown in \textbf{Fig \ref{overview}}, we have a difference neural network architecture for it. The vocal segmentation model is trained with a convolutional neural network \cite{cnn}. The vocal separation task is solved with our custom network based on U-Net architecture \cite{unet}. And the last one, we propose an Bidirectional LSTM network architecture designed to perform well at classification. 

\subsection{Vocal Segmentation}
The goal is to automatically detect such boundaries in audio signals so that the results are close to human annotation. In our problem, we applied Convolutional Neural Network (CNN) to predict the type of the music segments. We propose to train a neural network on human annotations to predict likely musical boundary locations in audio data. Our classifier is trained with an in-house collection of 2034 Karaoke songs. For every song in this dataset, the start point and endpoint of each vocal segment are annotated. We use this data set because it is easy to differentiate the vocal segments from the non-vocal ones from Karaoke songs; and these segments will be the training data for the vocal/non-vocal classifier. In our research, we used a CNN architecture combined with a few fully connected layers at the end. The overview of this network shown in \textbf{Fig \ref{cnn_figure}}. The input of model is 50 MFCC frames, each corresponding to a 500 milliseconds audio segment. The input frames are pass through two convolutional layers. First convolution layer have 128 filters with size \(10\times 10\) and the second layer have 32 filters with size \(5 \times 5\). After that, the output reduced by max-pooling layer \(5 \times 5\) and \(2 \times 2\). These layers are followed by two densely connected layers of 128 neurons, which are associated to dropout rates of 0.75 and 0.5. ReLU  activation function \cite{dahl2013improving} are used between each layers. The output layer is composed of two neurons, normalized using a softmax function to classify vocal and non-vocal class.

\subsection{Vocal Separation}
We first process the obtained vocal-present tracks before feeding it into our vocal separation network. The audio files are chopped into 6-second long snippets, then passed through the Short-time Fourier Transform (STFT) \cite{time-freq_feat_rep}. The short-time Fourier Transform features temporal frequency properties in every short timeframe. After obtaining the magnitude and phase matrices from the STFT, the magnitude matrix is then normalized to the logarithmic scale with preserving nonnegativity property using \texttt{log1p}:
\begin{equation}
\mathrm{log1p}(x)=\log(1+x),    
\end{equation}
intuitively, we are making a spectrogram similar to one in decibel scale - the normal one that is in everyday usage. The phase matrix, if needed, will be used to reconstruct the vocal track with the spectrogram output from our vocal separation model. 

\begin{figure}[t]
    \centering
    \includegraphics[width=8cm,height=4cm]{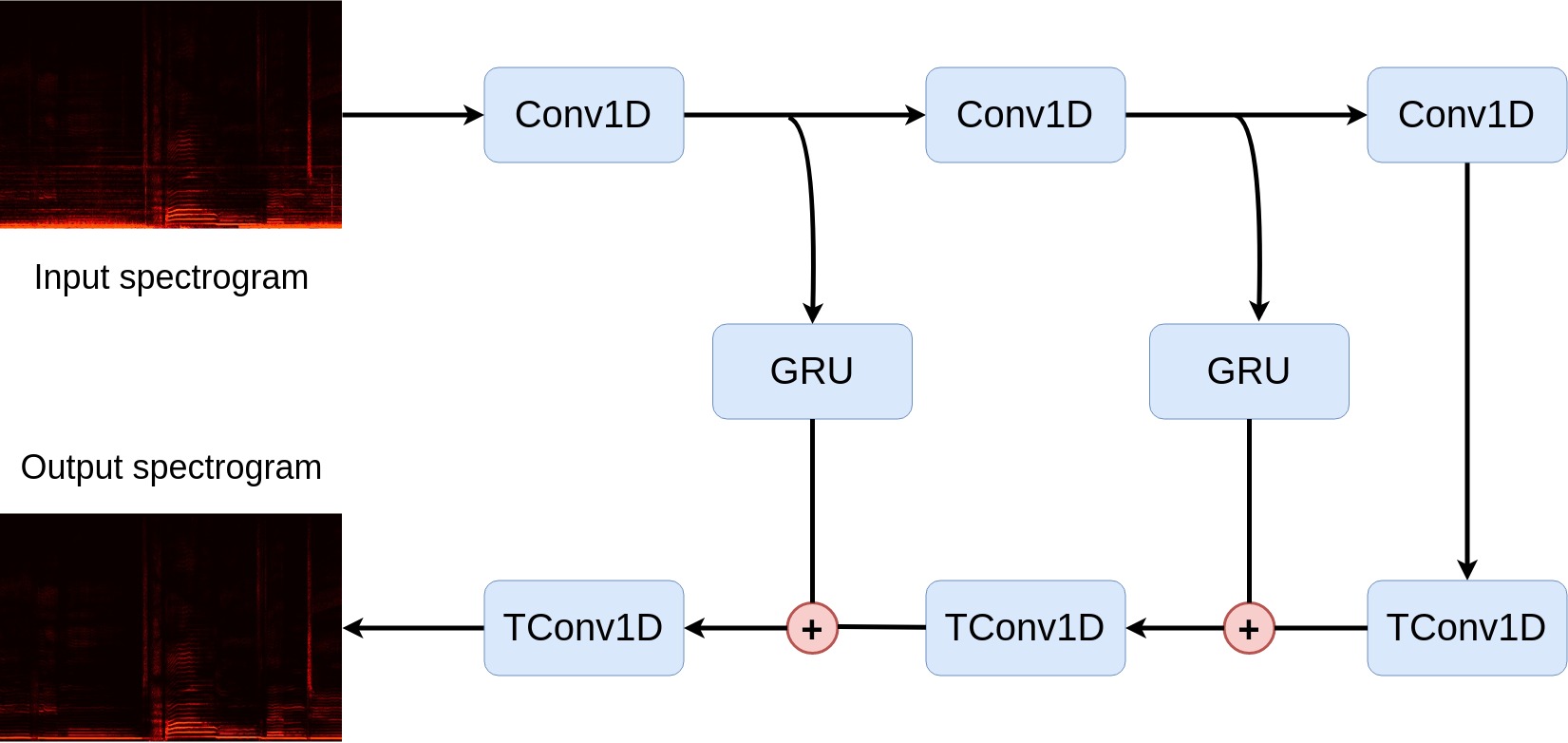}
    \caption{The vocal separation model architecture based on U-net with GRU skip-connection.}
    \label{voxtrac_model}
\end{figure}

Most current researches on vocal extraction uses pixel-wise segmentation on the spectrogram of the master track, then combines the result with the original phase matrix, to achieve the final result \cite{Roma2016SingingVS}. Notably, Jansson \textit{et. al} used a deep U-Net architecture for the task \cite{spotify-unet}, yielding decent result without blowing up the number of parameters. In light of such papers, we opted to use a similar but customized deep neural network described as follows. From the obtained spectrogram, we start with three 1-D convolution layers along the time domain for feature extractions. After that, there are three 1-D transposed convolution layers to expand and convert the features back into the same dimension as the output of its corresponding convolutional output. And last but not least, for the skip-connection layers, instead of just purely adding/concatenating the two encoded outputs from the convolutional layers into the input of the transposed-convolutional layers, we pass each said output through a Gated Recurrent Unit \cite{gru} layer. These layers will learn to map these matrices from the convolutional output spaces to the transposed convolutional input spaces, effectively bringing through information that have been lost in the downsizing process. The skip-connection is a design choice borrowed from the famous U-Net architecture; however we decided on having our little GRU twist since it makes little sense combining features of different latent spaces. Albeit working in practice, in our opinion, it would have made more sense if we reused the convolutional matrix in our transposed convolutional layer, which would decrease the number of tunable parameters, and the model's capacity as a result. The diagram \textbf{Fig \ref{voxtrac_model}} shows a visualization of the model.

The spectrogram from the last transposed convolutional layer, after undoing the \texttt{log1p} operation, will be the spectrogram of the extracted vocal track, and we can recover the actual track by passing that spectrogram and the phase matrix of the original track into the inverse STFT. However we will not be needing that, since only the spectrogram will be passed on to the next step.

\subsection{Vocal Classification}
With the vocal-only spectrogram obtained from the last step, we extract the Mel-Frequency Cepstral Coefficients (MFCCs) to be the features passed onto our model. The MFCCs, in contrast to a normal magnitude spectrogram, captures more detailed low-frequency features that correspond to human voice, while discarding the less informative part (the amount of information kept is a hyperparameter). Each frame, 13 MFC coefficients were extracted using 26 filter bands. To better model the behavior of the signal, the differentials and accelerations of the MFC coefficients were calculated. All these features were combined into a feature vector of size 39. The feature vectors served as input to the LSTM model. A Bidirectional-LSTM model with 3 hidden layers and 25 hidden units each layer is used for vocal classification. The result is then finally passed through a dense layer activated by the softmax function to get a predicted probability distribution of whether the track belongs to an artist.  Training was done using a backpropagation length of 20 time steps, with batch size of 64. The Adam optimizer with an initial learning rate of 0.001 is used to train this model.

\section{System setup and dataset}
\subsection{Dataset}
\begin{figure}[b]
    \centering
    \includegraphics[width=9.2cm,height=6cm]{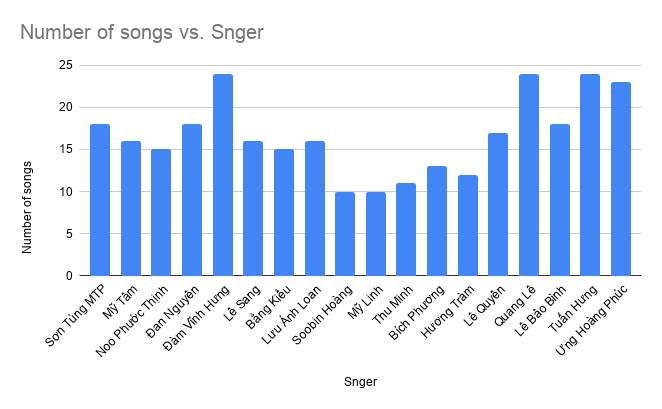}
    \caption{The number songs of each singer in our collected dataset.}
    \label{vn_song_dataset}
\end{figure}

In any machine learning and deep learning tasks, data plays an important role to the accuracy of the whole system. Thus, the data preparation phase must be carried out carefully. Unlike other problems, the singer classification problem is divided into 3 subproblems and each of them needs a different data set. Specifically, with the vocal segmentation problem, we use the 2034 Karaoke songs dataset to train the model. This dataset includes Vietnamese karaoke songs with annotations of vocal segments' starting and ending points. The vocal separation are trained on MUSDB18\footnote{https://sigsep.github.io/datasets/musdb.html}\cite{musdb18} and DSD100\footnote{https://sigsep.github.io/datasets/dsd100.html}\cite{dsd100} dataset. Two datasets are contain the full lengths music tracks of different styles along with their isolated drums, bass, vocals and other stems. To train model for the singer classification task, we collected 300 Vietnamese songs of 18 singers. Details of this data set are described in \textbf{Fig \ref{vn_song_dataset}} .

\subsection{System and hardware}
Our experiment is conducted on a computer with Intel Core i5-7500 CPU @3.4GHz, 32GB of RAM, GPU GeForce GTX 1080 Ti, and 1TB SSD Harddisk. All three subnetworks are implemented with the well-known PyTorch framework \cite{pytorch}.

\section{Experiments and results}
\subsection{Vocal and non-vocal segmentation result}
The raw frame-level vocal/non-vocal probabilities are obtained
using a step size of 10 milliseconds. The Viterbi \cite{forney1973viterbi} algorithm is used to infer the most likely voice segment (vocal or non-vocal) from these raw data, allowing an increase in the accuracy of our model. With the dataset of 2034 Karaoke songs, we use 1500 of those for training, and 534 for testing. We also divide this dataset into 4 music genres to analyze the output of our model. \textbf{Table \ref{vocal_segmentation_result}} shows the detail of our model. According to the result, we found that genres like country, ballad and bolero gave better classification results (approximately 98\%). This gives us a suggestion to improve our method in the next steps.


\begin{table}[t]
\centering
\caption{Vocal and non-vocal segmentation result}
\begin{tabular}{|l|l|l|l|l|l|l|}
\hline
\multirow{2}{*}{Song genre} & \multicolumn{3}{l|}{CNN Precision}                                  & \multicolumn{3}{l|}{CNN + Viterbi Precision}                        \\ \cline{2-7} 
                            & Vocal & \begin{tabular}[c]{@{}l@{}}Non\\ vocal\end{tabular} & Mean  & Vocal & \begin{tabular}[c]{@{}l@{}}Non\\ vocal\end{tabular} & Mean  \\ \hline
Country                     & 91.30 & 97.20                                               & 94.25 & 97.82 & 99.64                                               & 98.73 \\ \hline
Balad                       & 92.85 & 94.24                                               & 93.55 & 98.65 & 99.86                                               & 99.26 \\ \hline
Bolero                      & 94.32 & 90.24                                               & 92.28 & 96.30 & 98.12                                               & 97.21 \\ \hline
Rock                        & 88.23 & 97.15                                               & 90.69 & 90.64 & 90.67                                               & 97.10 \\ \hline
\end{tabular}
\label{vocal_segmentation_result}
\end{table}
As we can see, the precision for the first three genres was decent; however for the Rock genre it fell short. For future work, we would want to improve performance for bad-performance genres like it, as well as genres that are not in the current dataset (for e.g., electronic music, hip-hop, etc.)

\subsection{Vocal separation results on two datasets}
The evaluation is conducted by using the 
MUSDB18 dataset (100 songs for training and 50 songs for testing)
and the official packages from SiSEC2018 \cite{ward2018sisec}. We used the signal-to-distortion ratio (SDR) \cite{uhlich2017improving} \cite{liutkus20172016}, as it is the most widely used metric in this research problem. The model's evaluation with two datasets, DSD100 and MUSDB18, is introduced in Section 4. The detail of the result are shown in \textbf{Table \ref{vocal_separation_result}}.

\begin{table}[t]
\centering
\caption{The result of vocal separation}
\begin{tabular}{|l|l|l|}
\hline
                     & DSD100 & MUSDB18 \\ \hline
GRU Skip connection  & \textbf{5.92}   & 5.84    \\ \hline
LSTM Skip connection & 5.82   & 5.78    \\ \hline
\end{tabular}
\label{vocal_separation_result}
\end{table}

Audibly, the separated result is only decent enough for our task, but not on the production-level quality we hoped for. Currently we are experimenting with both improvements to the current model (adding attention, changing the skip connection layer), and other promising architectures. These considerations are however too specific for this paper, and will be further analyzed on some future paper on this sole subtask.

\subsection{Vocal classification result}
After vocal segmentation step, we do vocal classification experiments with MFCC from two audio signals. In the first experiment, we tried with raw features after concatenating the vocal segments. This audio signal includes vocals, instruments  and other sounds. The second experiment, we pass the raw signal audio through a vocal separation model. The output of this step is the input of our vocal classification model. Both experiments were performed with the same network architecture using Bidirectional LSTM with MFCC feature described in Section 2. The comparison of these experiments are shown in \textbf{Table \ref{vocal_classification_result}}. We use F1-score for this evaluation. 

\begin{table}[t]
\centering
\caption{The result of vocal classification with two audio signal}
\begin{tabular}{|l|l|l|l|}
\hline
                 & Mean precision & Mean recall & Mean F1 score \\ \hline
Raw signal       & 85.4           & 82.6        & 83.96         \\ \hline
Separated signal & \textbf{93.94}          & \textbf{91.78}       & \textbf{92.84}         \\ \hline
\end{tabular}
\label{vocal_classification_result}
\end{table}

Our goal leaves more to be desired. For starter, we can experiment with other features such as linear predictive coding, which is widely used in speaker recognition \cite{LPC}. Also, the current model can only handle songs with only one singer -- with little change, we can adapt this code to songs with multiple singers, given that each section of the song only has one singer. Another improvement we can add is adding multiple-singer detection in vocal mixes, say, when the voices are harmonizing, similar to speakers' sources separation. Further, we can experiment with singer embedding, given a singer's extracted features, we may be able to generalize about the vocal properties, the song style of that artist, etc, which is a hot topic in the music information retrieval community.
\section{Conclusion}
In this paper we have employed deep learning techniques to build neural networks to solve the singer vocal classification problem. We have proposed a method to solve this problem including the following steps: vocal segmentation, vocal extraction and vocal classification. Each of the steps above is addressed with the appropriate neural network architecture. This makes it easy for us to individually optimize each subproblem. The overall accuracy of this problem is approximately 93\% with the data set of 300 songs from Vietnamese singers. This dataset was also collected manually and publicly for the scientific community to conduct similar studies.

\begin{acks}
This work is partially supported by \textbf{\textit{Sun-Asterisk Inc}}. We would like to thank our colleagues at \textbf{\textit{Sun-Asterisk Inc}} for their advice and expertise. Without their support, this experiment would not have been accomplished.
\end{acks}

\bibliographystyle{ACM-Reference-Format}
\bibliography{ref}

\end{document}